\documentclass[runningheads]{llncs}
\usepackage[utf8]{inputenc}
\usepackage{graphicx}
\usepackage{svg}

\usepackage{comment}
\usepackage{todonotes}

\usepackage{booktabs}
\usepackage{multirow}

\begin{document}

\title{A Bibliometric Analysis of the BPM Conference Using Computational Data Analytics}
\titlerunning{A Bibliometric Analysis of the BPM Conference}
%
\author{Fabian Muff\orcidID{0000-0002-7283-6603} \and
Felix Härer\orcidID{0000-0002-2768-2342} \and
Hans-Georg Fill\orcidID{0000-0001-5076-5341}}
\authorrunning{F. Muff et al.}
%
\institute{University of Fribourg, Research Group Digitalization and Information Systems
\email{fabian.muff|felix.haerer|hans-georg.fill@unifr.ch}\\
\url{http://www.unifr.ch/inf/digits}}
\maketitle              
%

\begin{abstract}
The BPM conference has a long tradition as the premier venue for publishing research on business process management. For exploring the evolution of research topics, we present the findings from a computational bibliometric analysis of the BPM conference proceedings from the past 15 years. We used the publicly available DBLP dataset as a basis for the analysis, which we enriched with data from websites and databases of the relevant publishers. In addition to a detailed analysis of the publication results, we performed a content-based analysis of over 1,200 papers from the BPM conference and its workshops using Latent Dirichlet Allocation. This offers insights into historical developments in Business Process Management research and provides the community with potential future prospects.

\end{abstract}

\section{Introduction}

Whereas business process management is a topic that has been and is traditionally being treated by many different disciplines and outlets ranging from management sciences to computer science~\cite{VanAalst2013}, the International BPM Conference is today one of the most prestigious and well-recognized outlets for BPM research~\cite{ReckerM16}. Since its first edition in 2003, not only a large number of papers has been presented at the conference itself. In addition, multiple workshops have accompanied the conference, as well as various specialized forums such as the Blockchain or the Robotic Process Automation forum~\cite{bcforum2020}. Behind these papers stands a large international community with researchers from academia and industry. It can therefore serve as a good sample set for characterizing the discipline of business process management.

Despite previous analyses of the conference's papers, as for example conducted by Recker and Mendling~\cite{ReckerM16}, no computational analysis on the topics of the papers, the origin of their authors, the share of academic and industry contributions, or the evolution of topics over time has been performed recently. Previous analyses reverted mostly to subjective approaches, e.g.\ by using manual tagging of papers~\cite{VanAalst2013}, through specifically developed coding schemes~\cite{ReckerM16}, or restricted the analysis to previous time periods and the identification of trends~\cite{HouySFL13}.

We have therefore put together the following research questions in order to get further insights into the evolution of BPM research that have not been tackled before:

\begin{itemize}

    \item RQ 1: In which geographical regions is BPM research conducted?
    
    \item RQ 2: How has the quantity of papers at the BPM conference evolved over time, taking into account specific regions?
    
    \item RQ 3: How many authors are active in the BPM conference, what is their typical number of papers and do they work in academia or industry?
    
    \item RQ 4: What are the major topics at the BPM conference in terms of published research, how do they differ in terms of the main conference and workshops and how did they evolve over time?
    
    \item RQ 5: Is there a difference between topics covered by researchers in academia and such covered by researchers in industry?

    \item RQ 6: Is there an indication of prospective topics, application areas, or domains that have not yet been covered by the BPM conference but that are deemed relevant?

\end{itemize}

The method that we use for answering these questions is a computational, i.e.\ machine-based analysis of the metadata and contents of the papers at the BPM conference and its workshops. This permits us to avoid a coding bias and gives us transparent insights into the evolution of BPM research. The remainder of the paper is organized as follows: In Section 2 we discuss related analysis approaches, followed by a description of the research methodology in Section 3. Subsequently, we present the results from a descriptive and a content-based analysis of the papers in Sections 4 and 5. We then interpret the results of this analysis in Section 6 and conclude with limitations of the study and an outlook in Section 7.

\section{Related Work}
\label{relatedWork}

The investigation of literature sources for assessing the state-of-the-art of a discipline is a core part of any research endeavor and requires rigorous documentation~\cite{webster2002,brocke2009}. This process has and is still frequently conducted manually and with large personal effort. As a consequence, such analyses are often limited to a subset of available sources, e.g.\ by considering only particular outlets. For circumventing this limitation and enhancing the analysis, automated approaches have been proposed in addition~\cite{watson2020analysing,HouySFL13,HarerF20}. In the following we review the most recent previous, manual and automated literature analyses of the BPM discipline. 

In 2013, van der Aalst~\cite{VanAalst2013} conducted a comprehensive survey on BPM. His work presents twenty different use cases for BPM and how the main concerns of the BPM community are covered by them. For this purpose, 289 papers from the BPM conferences of 2003-2011 and of a previously published book were assigned to one or more use cases. In addition, each paper was manually tagged with one or more key concerns in BPM from a set of 342 tags. This permitted insights into the evolution of key concerns over time. The analyses in this paper are however based on a subjective assessment and assignment. 

In 2016 Recker and Mendling published an analysis of 347 papers published at the BPM conference~\cite{ReckerM16}. For their analysis, they focused on the identity and progress of the BPM conference. This included a classification study and an analysis of citation data. For this purpose, all papers from the conference proceedings from 2003 to 2014 were included. The classification of the papers followed a specifically-developed coding scheme that was applied by the authors in a manual process by reading the full-texts of the papers. The categories for the classification comprised the focus and intent of each paper, the research components, the research method, the positioning within the BPM lifecycle, the empirical evidence, and the type of implementation. For measuring impact, citation data was extracted for each paper using Google Scholar. 

An automated literature analysis method has been described by Houy et al.~\cite{HouySFL13} and applied to 905 papers in the Business Process Management Journal and the BPM conference proceedings for the years 2005-2011. Their work focuses primarily on past trends in the community. 
However, only the abstracts of the individual papers are analyzed and a thesaurus had to be derived by a domain expert in addition. Overall, the focus is put on individual terms and term groups, and the trends are derived over the entire period for individual terms. 

More recently, Neder et al. conducted an automated trend analysis in business process management~\cite{neder2018}. They reverted to a set of 661 papers extracted from the Web of Science database on the topic of business process management in the time frame 1995-2018. The papers were divided into those focused on business and management and those associated with information technology for comparing these two directions. With this data, set a semantic network analysis based on TF-IDF metrics was conducted for identifying the evolution of concepts over different time frames.

In summary, we can state that several profound analyses of BPM as a discipline have been conducted in the past based on the available literature. Apart from manual analyses, some sources also report on automated, computational approaches. What is however missing is a holistic computational approach joining descriptive and content-based metrics that are verifiable.



\section{Research Methodology}
\label{researchMethodology}

Unlike most previous studies and surveys in the BPM community, the analysis methods in this work are two-fold. After automated data collection and semi-automated cleaning steps, an ETL workflow with multi-dimensional analysis~\cite{AliW17} is applied at first for descriptive statistics. Secondly, existing data is enriched with full texts for applying bibliometric topic identification methods. For classifying the process as whole, it follows the well-known data mining and data analysis approach KDD~\cite{FayyadPS96} in its process of data selection, preprocessing, transformation, data mining, and interpretation and evaluation. Our method is illustrated in Fig. \ref{fig:DCAP} and further explained in Section \ref{DataCollection}.

\subsection{Aims and Scope of the Study}
The aim of this work is to apply and validate a technologically rather new approach involving automatic data collection and bibliometric analyses to the discipline of BPM. As a first step, a quantitative analysis of the metadata of the BPM Conference and the BPM Workshops between the years 2005 and 2019 has been performed. By conducting and interpreting these analyses, we will highlight several aspects that characterize the BPM community. These include information on the geographical regions in which BPM research is conducted, the development of the quantity of papers and authors in the community, or the academic or industrial background of the participating institutions. Furthermore, we will show the different topics in relation to the conference and the workshops and analyze their evolution over time.

\subsection{Data Collection} \label{DataCollection}
A dump of the DBLP database from 2020-11-19 
 was used as the basis for data collection. The BPM conference and its workshops were chosen as the starting point. The XML file from the DBLP contained a total of 1,399 entries on BPM and BPM workshops respectively. In the vast majority of cases, these entries contained title, authors, year, outlet, URL, and DOI and were combined into a JSON file. In addition, this data was enriched using DOI.org and the publisher websites from Springer, adding possibly the DOI, affiliation and the country to the metadata. The data on publications and queries are publicly available~\cite{index_and_queries}.

\begin{figure}[htp]
    \centering
    \includegraphics[width=1\columnwidth]{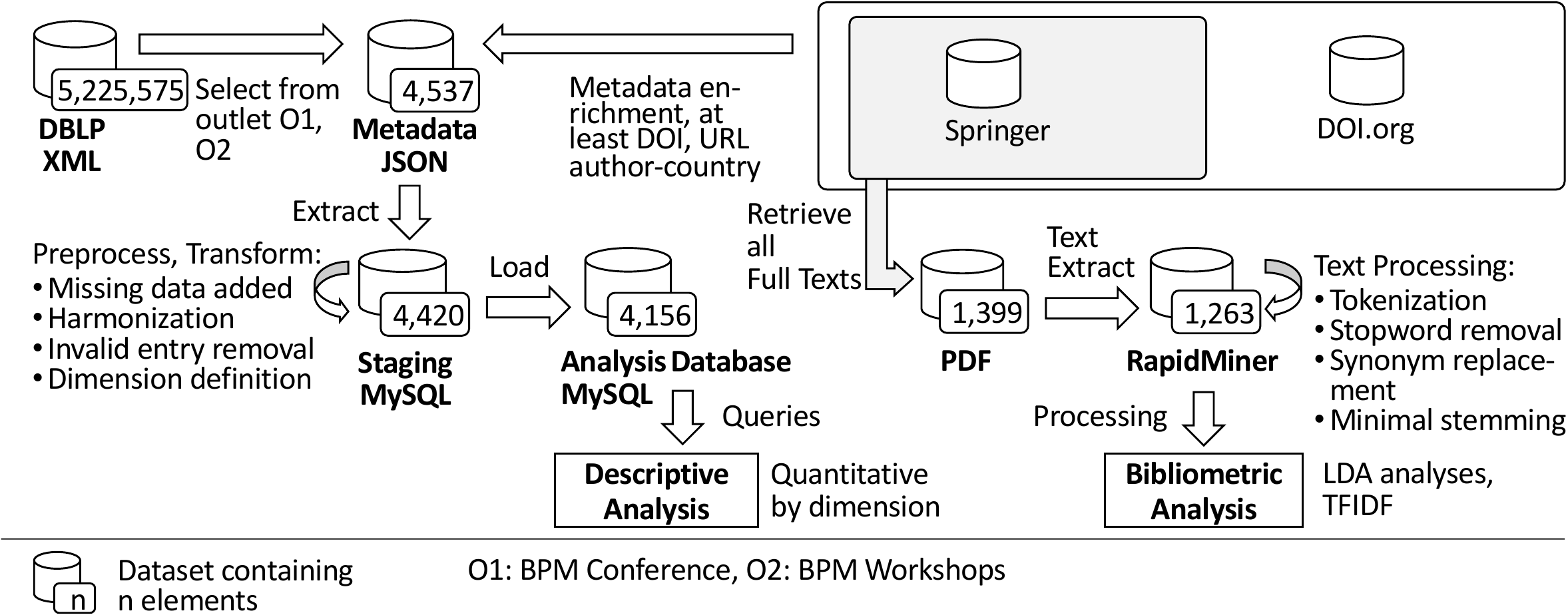}
    \caption{Data collection and analysis process based on the DBLP XML dataset}
    \label{fig:DCAP}
\end{figure}

%

After the collection of the raw data, the metadata was extracted into a staging database consisting of 4,420 entries. Then, a manual harmonization of all names, countries, institutions, cities and outlets, as well as the elimination of invalid entries, including non-paper posts, such as editorials and introductions and placeholders with missing authors, was conducted on this dataset. The remaining 4,156 entries were converted into the star scheme with partially normalized dimensions or "snowflaking" shown in Fig. \ref{fig:DCAP} for the purpose of analysis of multiple dimensions.

\begin{figure}[htp]
  \centering
  \includegraphics[width=1\columnwidth]{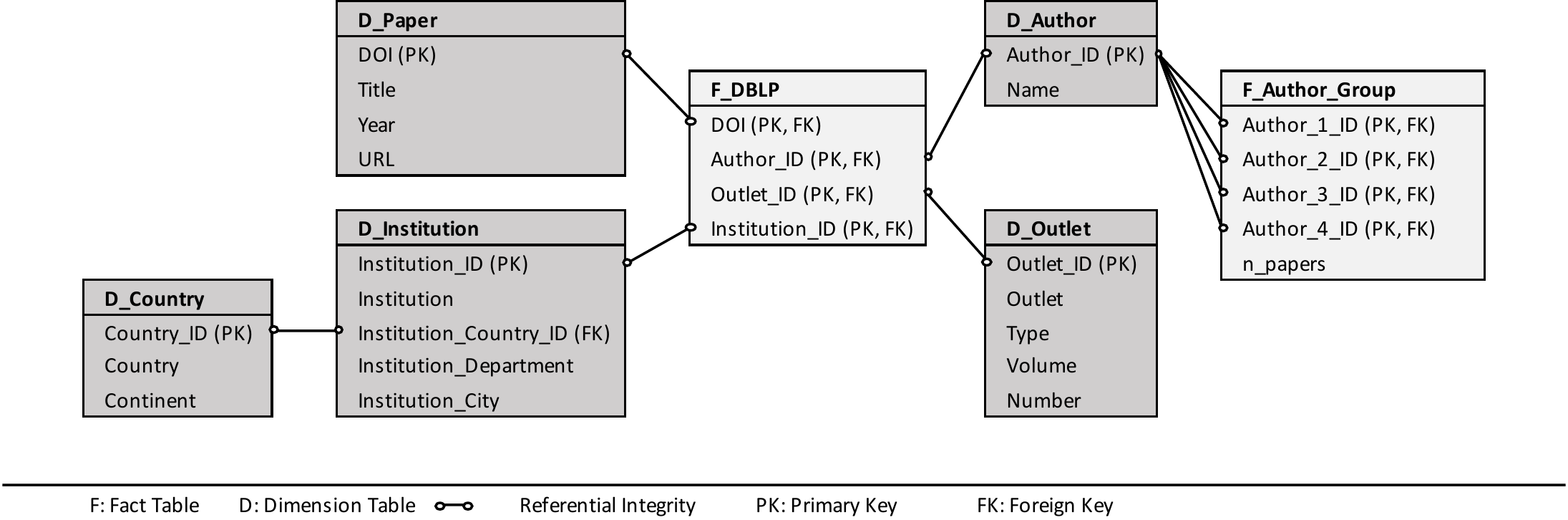}
  \caption{Schema of the analysis database. Fact tables (prefix F) store DBLP
publications and author groups according to dimension tables (prefix D)~\cite{HarerF20}. Note: for all author groups metrics are calculated; author IDs are only stored for up to the first four authors.}
    \label{fig:snowflake}
\end{figure}

Fortunately, we were able to rely on a single publisher for the full texts of the papers. The full texts could thus be downloaded automatically using a Node.js scraping script. Subsequently, the reduced set of 1,263 full texts, resulting from removing introductions and prefaces, was converted into text files and the titles of the respective papers were filtered out of the full texts in order to prevent a possible bias of the data basis. After that, the documents were loaded into RapidMiner Studio 9.8. Within RapidMiner the normalization of plurals and frequent inflected forms, as well as further NLP activities such as tokenization, stop word elimination, synonym substitution, and limited stemming were applied. 


\subsection{Data Analysis}
\label{dataAnalysis}

For the descriptive analysis, the database served as a direct source. Through a quantitative analysis, detailed results were retrieved by executing queries over the dimensions, e.g.\ for the frequencies of publications. For multi-dimensional queries, the database design proved useful, e.g.\ authors associated with organizations from countries with varying granularity levels, such as individual countries, continents or all countries. Second, a bibliometric data analysis was conducted on the full texts of the documents. For this purpose the Latent Dirichlet Allocation (LDA) method was used, which is a statistical tool for identifying topics in documents. Last, a TF-IDF analysis for unique terms occurring in the different topics was performed to get an overview of the evolution of these terms over time. 

\section{Descriptive Analysis of BPM Publications}
\label{descriptiveAnalysis}
The first descriptive analysis of the dataset targets the number of published papers and the number of publishing authors in each time period -- this is depicted in Table~\ref{Evolution over time on continents}. On a second axis, the different continents are shown as a further dimension. For the following analyses, the time period between 2005 and 2019 is considered. This is due to the fact that the BPM workshops have only been held regularly since 2005 and that the papers from 2020 were not yet available at the time of the analysis. In order to achieve a well-comparable data basis, the total period has additionally been divided into three intervals of five years each. 

\begin{table}
\centering
\caption{Evolution of the number of published papers (n pap.) and the number of authors (n auth.) over three time periods on different continents with sum and distinct total for showing overlapping values -- see SQL queries Q1-Q4~\cite{index_and_queries}.}
\label{Evolution over time on continents}
\begin{tabular}{l|c|c|c|c|c|c|c|c}
\hline
{\multirow{3}{*}{Continent}} & \multicolumn{8}{c}{Year}                                                                                                          \\ 
\cline{2-9}
\multicolumn{1}{c|}{}                           & \multicolumn{2}{c|}{2005 - 2009} & \multicolumn{2}{c|}{2010 - 2014} & \multicolumn{2}{c|}{2015 - 2019} & \multicolumn{2}{c}{2005 - 2019}  \\ 
\cline{2-9}
\multicolumn{1}{c|}{}                           & n pap. & n auth.           & n pap. & n auth.           & n pap. & n auth.           & n pap. & n auth.           \\ 
\hline
Africa                                          & 2        & 3                   & 11       & 21                  & 5        & 12                  & 18       & 35                  \\
Asia                                            & 32       & 65                  & 53       & 108                 & 36       & 74                  & 121      & 224                 \\
Europe                                          & 316      & 577                 & 365      & 647                 & 310      & 556                 & 991      & 1,512                \\
North America                                   & 46       & 87                  & 53       & 97                  & 23       & 32                  & 122      & 186                 \\
Oceania                                         & 63       & 78                  & 38       & 51                  & 35       & 60                  & 136      & 152                 \\
South America                                   & 10       & 29                  & 19       & 50                  & 36       & 65                  & 65       & 127                 \\ 
\hline\hline
\(\Sigma \)                                                  & 469      & 839                 & 539      & 974                 & 445      & 799                 & 1,453     & 2,236                \\
Distinct Total                                  & 417      & 820                 & 469      & 958                 & 377      & 782                 & 1,263     & 2,174               
\end{tabular}

\end{table}

Concerning the geographical regions of BPM research (RQ 1), it could be found, that authors from 35 countries authored papers at the conference and from 49 countries at the workshops. Thereby, Europe plays a major role, with 78\% of all articles having at least one author who was affiliated with a European institution at the time of publication. This applies to all three time periods. This insight can be further substantiated by regarding the number of papers and authors per country as shown in Table \ref{tab:authors_countries}. It shows that among the top ten countries in terms of published papers, eight are European countries and that these top ten countries contributed to 47\% of all published papers.

Regarding the evolution of the number of papers (RQ 2), a positive trend can be observed. This can be shown by a linear regression on the number of publications over time with \(F(1,13)=5.93, p=.029, R^2_{adjusted}=.26\). The regression coefficient \((B=1.8)\) indicates an average increase of 1.8 publications per year. By coming back to Table~\ref{Evolution over time on continents} again, we can find that there is no obvious continuous increase in publications except in South America. There, the average increase comprises 0.4 publications per year with \(F(1,11)=7.40, p=.020, R^2_{adjusted}=.35\).

\begin{table}
\centering
\caption{Number of papers and authors for the top ten countries in terms of the total number of papers. As authors may be affiliated to multiple institutions in different countries there is a difference in sum and distinct total-- see SQL queries Q4-Q5~\cite{index_and_queries}.}
\label{tab:authors_countries}
\begin{tabular}{l|c|c} 
\hline
Country        & n papers & n authors  \\ 
\hline
Germany        & 379      & 507        \\
Netherlands    & 194      & 141        \\
Australia      & 135      & 150        \\
Italy          & 122      & 194        \\
Austria        & 118      & 128        \\
United States  & 99       & 155        \\
Spain          & 98       & 112        \\
Belgium        & 61       & 65         \\
France         & 58       & 95         \\
United Kingdom & 43       & 65         \\ 
\hline\hline
\(\Sigma \)            & 1,307     & 1,612       \\
Distinct Total & 1,031     & 1,564      
\end{tabular}
\end{table}           

In RQ 3 we considered how many authors are active in the BPM conference and its workshops, their typical number of papers and their institutional affiliation. As shown in Table \ref{avg_participations}, the number of distinct authors at the BPM workshops amounts to 1,655 authors, which is more than twice as high as at the conference. This is most likely due to the fact that almost twice as many papers were published at the workshops as at the conference itself. By regarding the average number of authors per paper, we note that this value amounts to 3.2 for the conference and 3.3 for the workshops over the entire time period. This suggests that there is no fundamental difference between the conference and workshops. However, when regarding the development of the average number of papers per author over the different years, we see that this number increases steadily, as shown by the total values in Table~\ref{avg_participations}. Note that this row is a distinct total since some authors published papers at the conference and the workshops. The increasing trend can again be confirmed by a regression analysis with the number of authors per paper with \(F(1,13)=21.26, p=.0005, R^2_{adjusted}=.62\) and a regression coefficient of $B=.05$, indicating an average yearly increase in the number of authors per paper by .05 over all years.


\begin{table}
\centering
\caption{Number of papers published at the conference and at workshops in the three time periods with the total (n authors) and average number of authors per paper (avg authors/paper), as well as the average number of papers per author (avg paper/author) -- see SQL queries Q6-Q8~\cite{index_and_queries}.}
\label{avg_participations}
\begin{tabular}{l|l|c|c|c|c} 
\hline
                                & Period            & 2005 - 2009 & 2010 - 2014 & 2015 - 2019 & 2005 - 2019  \\ 
\cline{2-6}
\multirow{3}{*}{Conference}     & n authors         & 363         & 307         & 282         & 803          \\
                                & avg authors/paper & 2.9         & 3.3         & 3.5         & 3.2          \\
                                & avg paper/author  & 1.3         & 1.5         & 1.6         & 1.7          \\ 
\hline
\multirow{3}{*}{Workshops}      & n authors         & 555         & 852         & 597         & 1,655         \\
                                & avg authors/paper & 3.0         & 3.2         & 3.5         & 3.3          \\
                                & avg paper/author  & 1.3         & 1.4         & 1.4         & 1.6          \\ 
\hline
\multirow{3}{*}{Distinct Total} & n authors         & 820         & 958         & 782         & 2,174         \\
                                & avg authors/paper & 3.0         & 3.3         & 3.6         & 3.3          \\
                                & avg paper/author  & 1.5         & 1.6         & 1.7         & 1.9         
\end{tabular}
\end{table}




A slight increase can be observed in the average number of papers authored or co-authored by an author over the three time periods for the conference and the workshops. In addition, the average number of authors per paper has also increased over time. Over the entire time period, the average number of papers per author amounts to 1.91 for all papers from the conference and workshops. By analyzing the individual years, a weak increase in the average number of papers per author over time can be confirmed by a regression analysis. This shows an average yearly increase in the number of papers authored or co-authored by the same author of $B=.006$ with \(F(1,13)=2.86, p=.11, R^2_{adjusted}=.18\).

In Table~\ref{institution_categories}, the number of authors and institutions according to affiliation categories is shown. The institutions were divided into four categories. These comprise purely academic ones such as universities and universities of applied sciences (\textit{Higher  Level  Education  Institutions}), institutions with academic and industrial relations such as Fraunhofer in Germany or Unité Mixte de Recherche (UMR) in France (\textit{Research Institutions}), purely industrial institutions, i.e.\ companies (\textit{Industry}), to organizations that could not be assigned to one of these categories (\textit{Other}), e.g.\ municipalities. The institutions were assigned to one of these categories manually~\cite{index_and_queries}. 

As shown in Table \ref{institution_categories}, an author is assigned to a given category if they co-authored at least one paper representing an institution that was assigned to this category. It is evident that most authors come from \textit{Higher Level Education Institutions} and that 388 out of 619 (62\%)  of all institutions observed in the dataset are assigned to this category. Another considerable proportion of authors work in \textit{Research Institutions} or in \textit{Industry}. It should be noted that the number of authors in the field of \textit{Research Institutions} is twice as high as in \textit{Industry}, with roughly the same number of institutions in each category. Furthermore, it must also be considered here that an author can be affiliated to multiple institutions. Therefore, a sum row and a distinct total row have been added to Table \ref{institution_categories}. This is to show these overlaps.


\begin{table}
\centering
\caption{Overview over the institution categories and the number of authors affiliated in these categories with sum and distinct total to show overlaps. The categories are \textit{Higher Level Education Institutions}, \textit{Research Institutions}, \textit{Industry} and \textit{Other} -- see SQL queries Q9-Q10~\cite{index_and_queries}.}
\label{institution_categories}
\begin{tabular}{l|c|c|c|c|c} 
\hline
\multirow{3}{*}{Category}                                                    & \multicolumn{5}{c}{Year}                                                   \\ 
\cline{2-6}
                                                                             & 2005 - 2009 & 2010 - 2014 & 2015 - 2019 & \multicolumn{2}{c}{2005 - 2019}  \\ 
\cline{2-6}
                                                                             & n authors   & n authors   & n authors   & n authors & n institution        \\ 
\hline
\begin{tabular}[c]{@{}l@{}}Higher Level Educ.\end{tabular} & 700         & 671         & 623         & 1,594      & 388                  \\
Research Institution                                                               & 228         & 216         & 121         & 435       & 106                  \\
Industry                                                                     & 94          & 93          & 40          & 209       & 110                  \\
Other                                                                        & 5           & 3           & 12          & 20        & 15                   \\ 
\hline\hline
\(\Sigma \)                                                                               & 1,027        & 983         & 796         & 2,258      & 619                  \\
Distinct Total                                                               & 820         & 958         & 782         & 2,174      & 619                 
\end{tabular}
\end{table}

\section{Content-based Analysis of BPM Publications}
\label{contentBasedAnalysis}

We examined the contents of the papers found in our dataset in the second phase. For this, we used the MALLET (\textbf{MA}chine \textbf{L}earning for \textbf{L}anguag\textbf{E} \textbf{T}oolkit) and the LDA implementation that is part of RapidMiner 9.8. LDA is a topic modeling methodology operating at the level of documents in order to classify their topics. In comparison to simpler approaches such as word frequency, TF-IDF, and n-gram analysis, LDA constructs a probabilistic model allowing for multiple topics per document. Given a set of documents, any document $d$ is represented by a statistical distribution \(\theta _{d} \) over its topics. That is, each subject has a specific probability or weight for \(d\), and for any topic \(k\) a distribution of words \(\theta _{d,k} \) \cite{Blei2012}. The hidden variables of the distributions are computed with parallel processing by the Gibbs sampling scheme, where per-word weights are determined so that their probability of occurring in a specific topic is maximized \cite{newman2009distributed}. 
For all conducted LDA studies we present the top five terms according to their weight (cf. \cite{Blei2012,rosen2004author}). The topics are sorted by cumulative weight, with the weight of a topic \(k\) and word \(w\) as occurrence measure over \(w\) assigned to \(k\). Note that it is only possible to include identified weights of the top five words. This procedure replicates closely the approach established before in \cite{HarerF20}. 

The bibliometric analysis with LDA was conducted on the full-texts of documents between 2005 and 2019 and the different sub-periods of all papers, over the workshop and conference category, over all continents, over a subset of institution categories, and over the most involved countries -- see~\cite{index_and_queries} for configuration. Due to space limitations we present in the following only a subset of the results.


\begin{table}[h]
\centering
\caption{LDA topics for all papers of the conference and the workshops from 2005 to 2019 ordered by cumulative topic weight.}
\includegraphics[width=.9\linewidth]{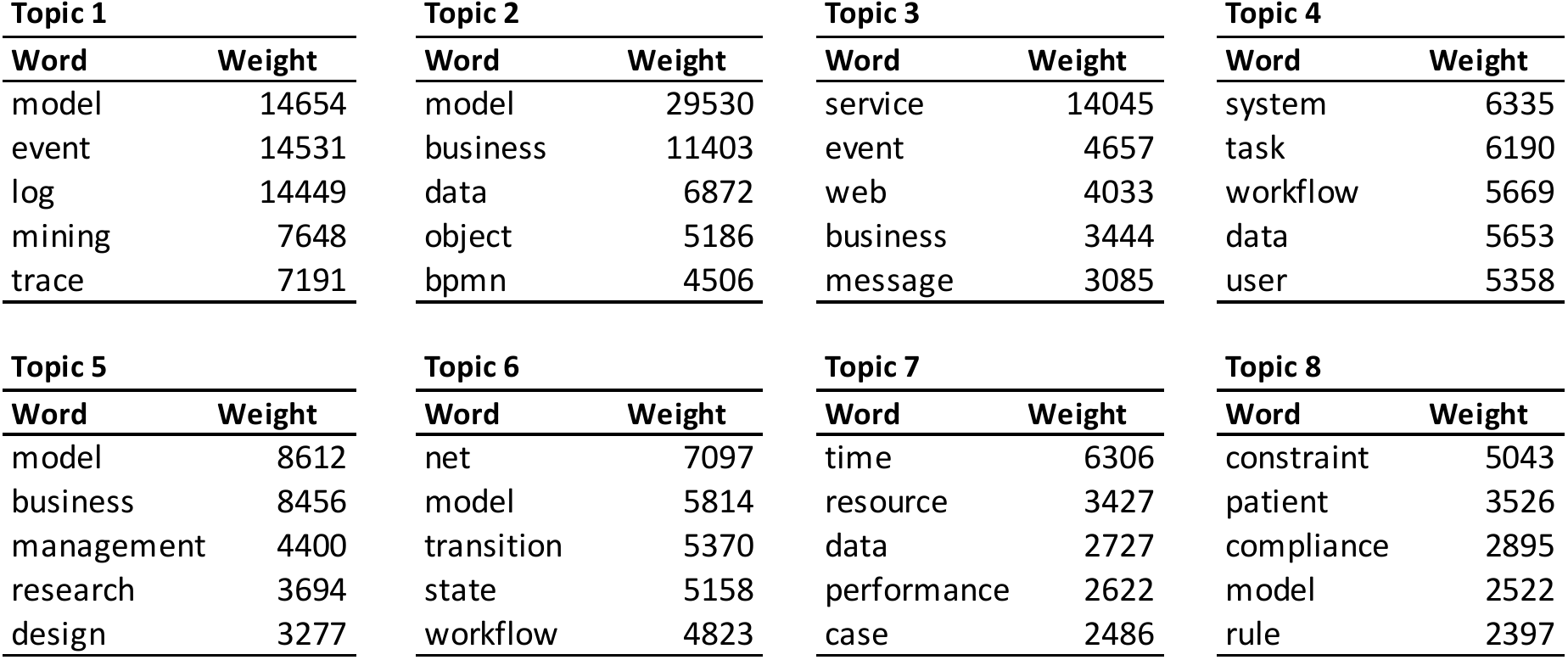}
\label{LDA topics for all papers from 2005 to 2019 ordered by cumulative topic weight}
\end{table}
\begin{table}[h]
\centering
\caption{LDA topics for all papers of the conference and the workshops from 2005 to 2009 ordered by cumulative topic weight}
\includegraphics[width=.9\linewidth]{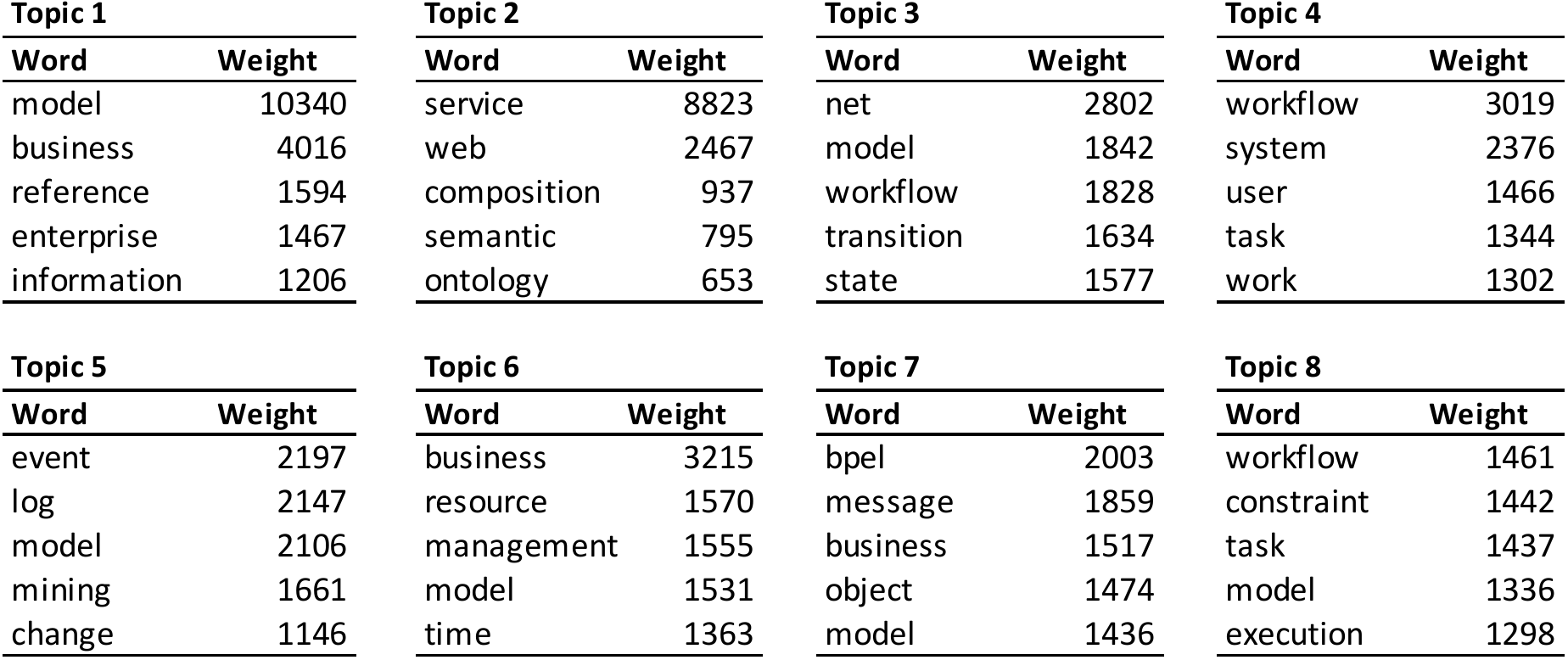}
\label{LDA topics for all papers from 2005 to 2009 ordered by cumulative topic weight}
\end{table}

\begin{table}[h]
\centering
\caption{LDA topics for all papers of the conference and the workshops from 2010 to 2014 ordered by cumulative topic weight.}
\includegraphics[width=.9\linewidth]{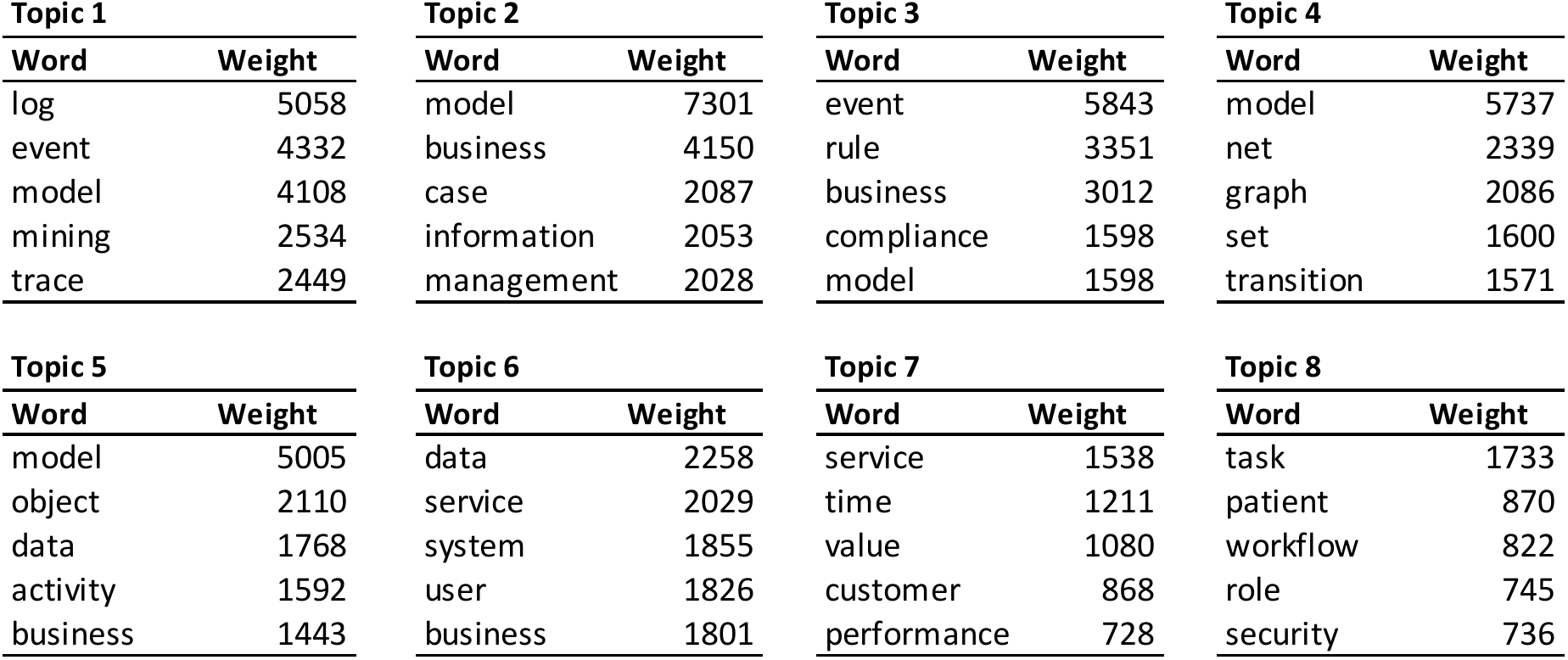}
\label{LDA topics for all papers from 2010 to 2014 ordered by cumulative topic weight}
\end{table}

\begin{table}[h]
\centering
\caption{LDA topics for all papers of the conference and the workshops from 2015 to 2019 ordered by cumulative topic weight.}
\includegraphics[width=.9\linewidth]{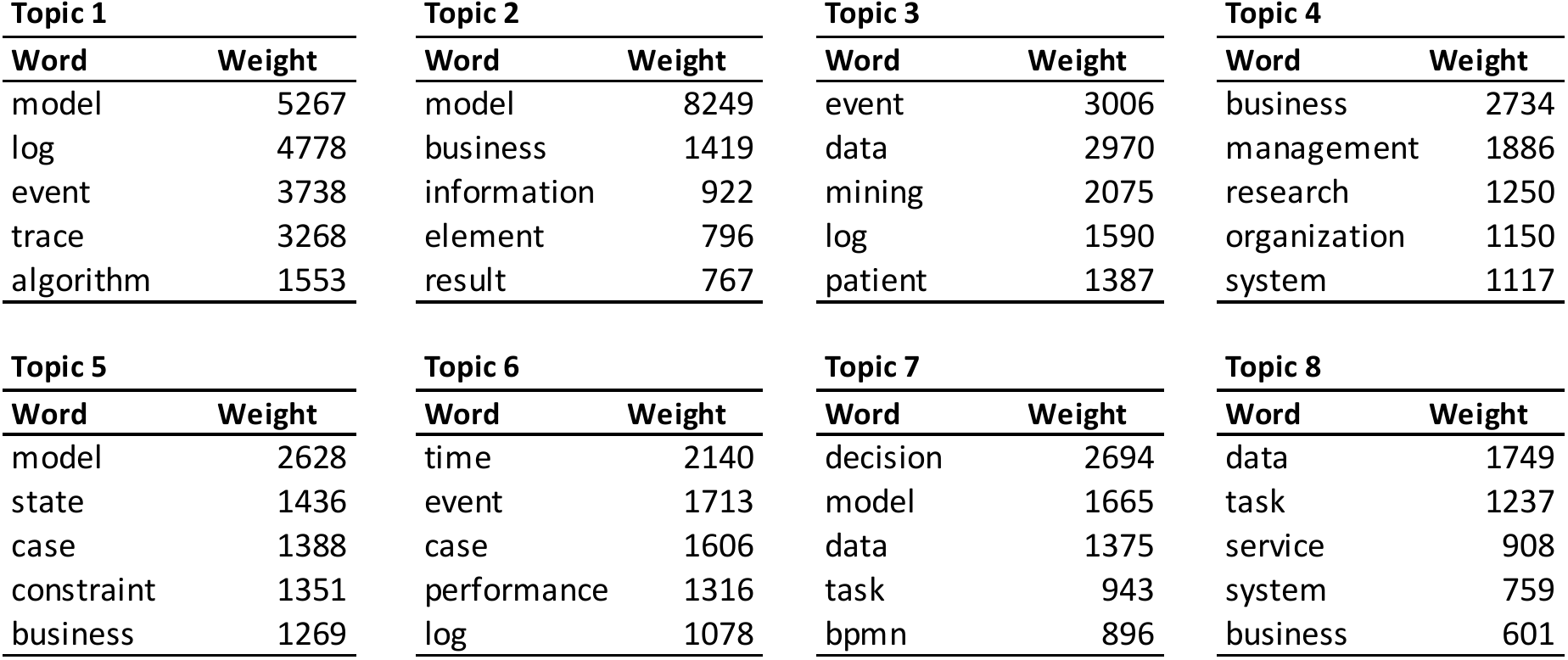}
\label{LDA topics for all papers from 2015 to 2019 ordered by cumulative topic weight}
\end{table}

The LDA results are shown in a standardized format. Per analysis, the top 8 topics are listed in a table, sorted by cumulative weight of the 5 included terms. Thus, a table always refers to an independent analysis and the weights of the individual terms can only be compared within a table. Table \ref{LDA topics for all papers from 2005 to 2019 ordered by cumulative topic weight} represents the results of the LDA analysis across all papers, i.e., the conference and workshops over the entire period from 2005 to 2019. Tables \ref{LDA topics for all papers from 2005 to 2009 ordered by cumulative topic weight}, \ref{LDA topics for all papers from 2010 to 2014 ordered by cumulative topic weight}, and \ref{LDA topics for all papers from 2015 to 2019 ordered by cumulative topic weight} are the results of the same papers, but only from the individual periods 2005-2009, 2010-2014, and 2015-2019. We can observe, for example, over all papers in the span 2015-2019, the term \textit{business} is prominent in the topics 2, 3, and 5. Topic 1 is the most weighted topic with the terms \textit{model}, \textit{event}, \textit{log}, \textit{mining}, and \textit{trace}. Topic 2 follows, including terms on \textit{business}, \textit{model}, and \textit{bpmn}, in addition to \textit{data} and \textit{object}. Further, we note in topic 3 the terms \textit{service}, \textit{event}, and \textit{web}, with topic 4 involving \textit{workflow} or \textit{system}.

\begin{table}[h]
\centering
\caption{LDA topics for conference papers only from 2005 to 2019 ordered by cumulative topic weight.}
\includegraphics[width=.9\linewidth]{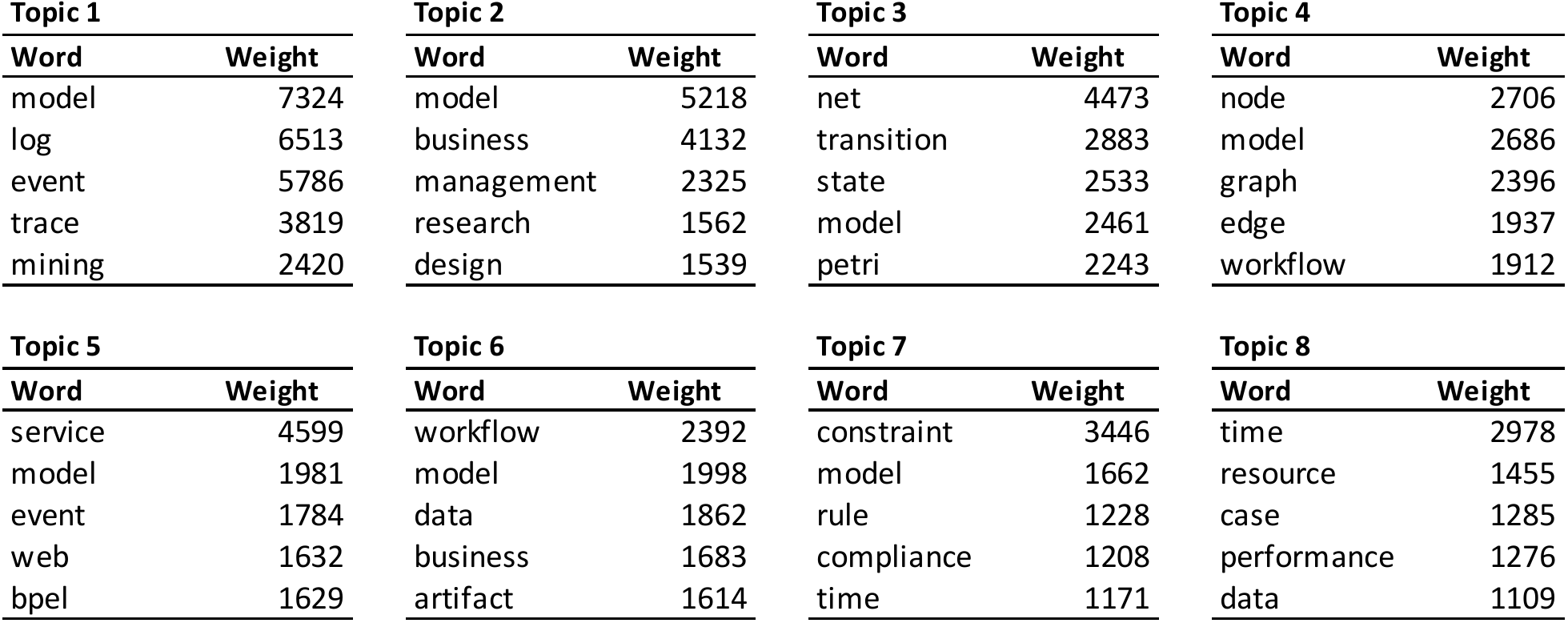}
\label{LDA topics for conference papers from 2005 to 2019 ordered by cumulative topic weight}
\end{table}
\begin{table}[h]
\centering
\caption{LDA topics for workshop papers only from 2005 to 2019 ordered by cumulative topic weight.}
\includegraphics[width=.9\linewidth]{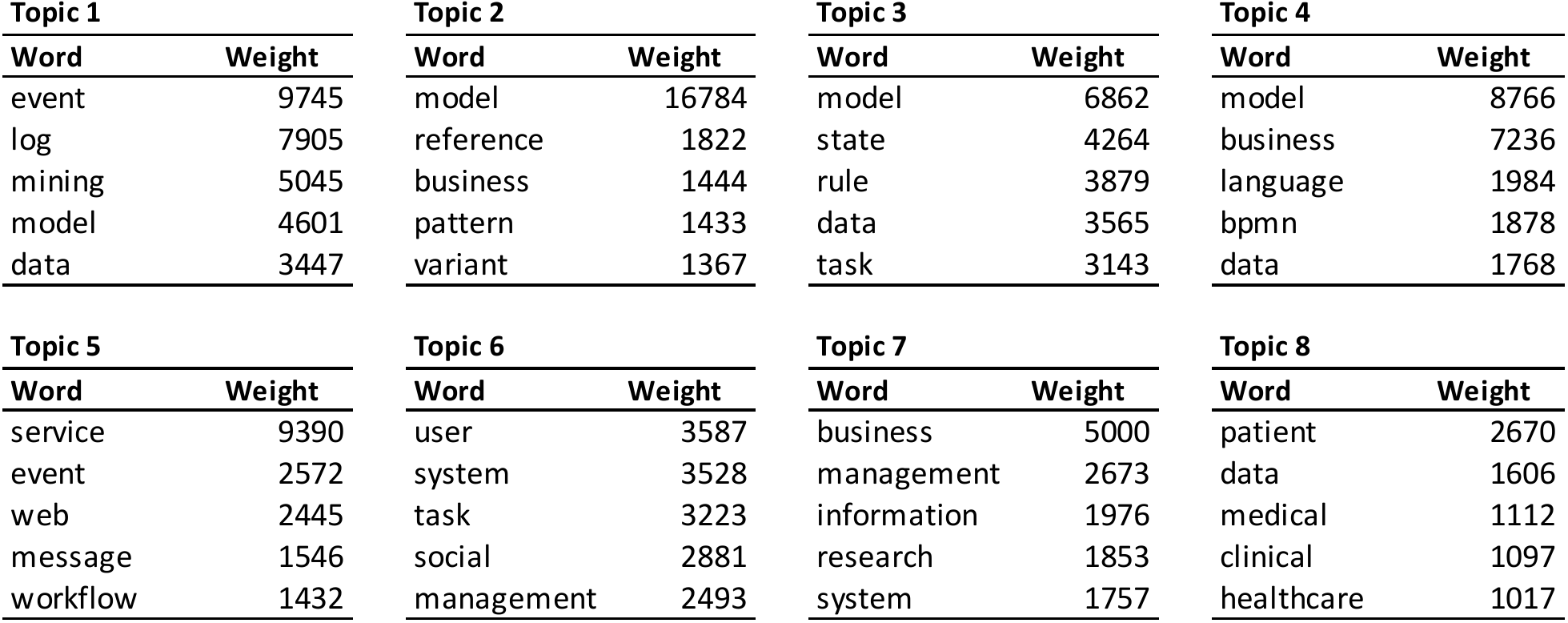}
\label{LDA topics for workshop papers from 2005 to 2019 ordered by cumulative topic weight}
\end{table}

Excerpts of the LDA analysis results for all conference papers from 2005 to 2019 are shown in Table \ref{LDA topics for conference papers from 2005 to 2019 ordered by cumulative topic weight}, respectively in Table \ref{LDA topics for workshop papers from 2005 to 2019 ordered by cumulative topic weight} for the papers published in workshops. In the conference and the workshops, prominent terms are \emph{mining} or \emph{workflow}, whereas \emph{patient} or \emph{healthcare} are only present in topics of workshops.

\begin{table}[h]
\centering
\caption{LDA topics for all papers of the conference and the workshops from 2005 to 2019 with institution category \emph{Higher Level Education} or \emph{Research Institution} ordered by cumulative topic weight.}
\includegraphics[width=.9\linewidth]{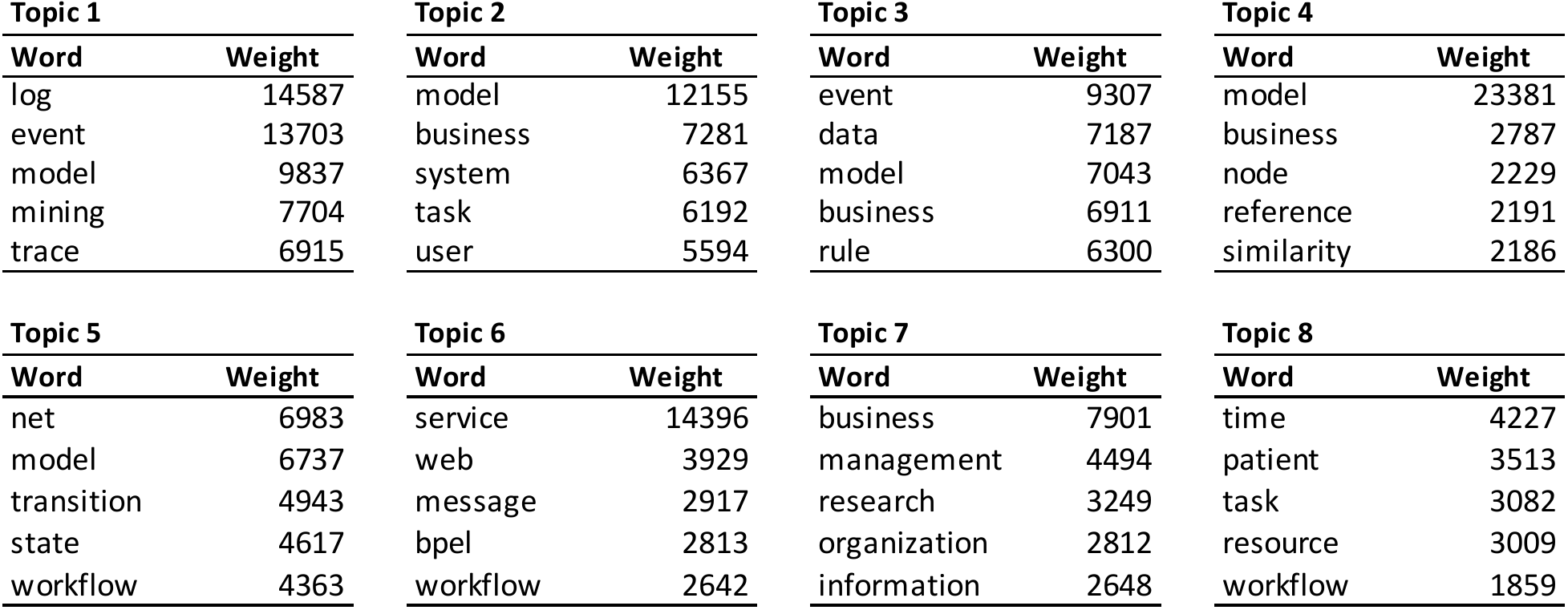}
\label{LDA topics for all papers from 2005 to 2019 with institution category 0 or 1 ordered by cumulative topic weight}
\end{table}
\begin{table}[h]
\centering
\caption{LDA topics for all papers of the conference and the workshops from 2005 to 2019 with institution category \emph{Industry} or \emph{Other} ordered by cumulative topic weight.}
\includegraphics[width=.9\linewidth]{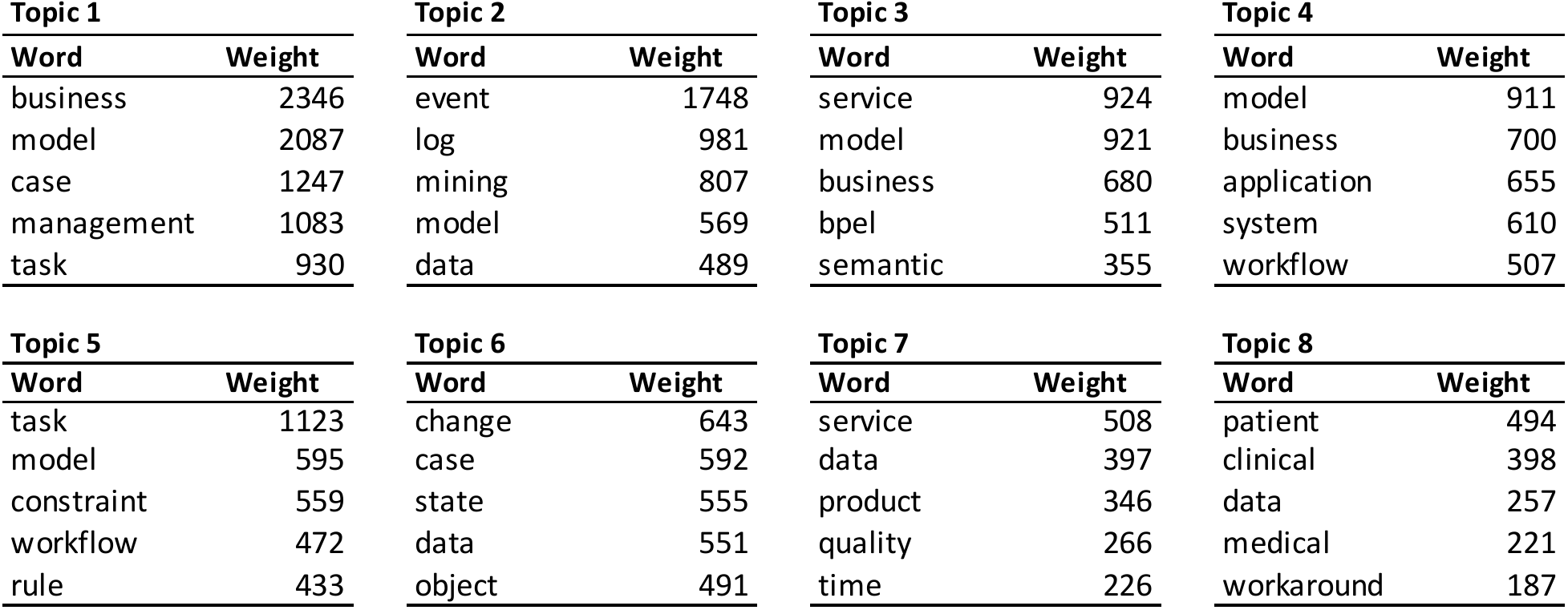}
\label{LDA topics for all papers from 2005 to 2019 with institution category 2 or 3 ordered by cumulative topic weight}
\end{table}

In Table \ref{LDA topics for all papers from 2005 to 2019 with institution category 0 or 1 ordered by cumulative topic weight} and \ref{LDA topics for all papers from 2005 to 2019 with institution category 2 or 3 ordered by cumulative topic weight} the results for the combination of the institution categories \textit{Higher Level Education Institutions} and \textit{Research Institutions}, resp. the categories \textit{Industry} and \textit{Other} are presented. This division was chosen since, according to our interpretation, the first two categories likely represent the academic and the second two categories the industrial sector. Here we can observe, for example, the terms \textit{petri} and \textit{node} for the academic sector only, whereas the terms \textit{clinical} and \textit{medical} are present in the industrial sector only.

\section{Discussion} \label{Discussion}
\label{discussion}

This section summarizes our main findings for the descriptive and content-based analyses and discusses possible interpretations in the context of business process management. Finally, we reflect on the identified research topics in light of prior work that used comparable methods by pointing out the topics supported by and differing from this study.


The research questions set out initially fall into two categories. RQ 1 to 3 can be answered directly in terms of plain data and descriptive statistics (Section \ref{descriptiveAnalysis}), related to geographical distribution (RQ 1), publications over time (RQ 2), and authors with their affiliations (RQ 3). The second subset of research questions RQ 4 to 6 involves the data and interpretation of the content-based analysis (Section \ref{contentBasedAnalysis}).

For RQ 1, the ample geographical diversity of BPM becomes obvious through the fact that people from all continents authored papers at BPM in the past. However, we can also note a strong European influence, and in particular of Germany and the Netherlands -- see~Tables~\ref{Evolution over time on continents}, \ref{tab:authors_countries}. 

Concerning the evolution of publications over time (RQ 2), the plain data indicates a slight upward trend in publications; however, the overall quantity is not representative of general research output since it is also determined by external factors, such as time and space restrictions of conference programs and proceedings. When adjusting for relative changes over time, the parameter of geographical origin reveals a remarkably stable participation share from Europe between 68\% and 69\% of all papers over 15 years, while South America is the only continent with a clear upward trend -- see Table~\ref{Evolution over time on continents}. 

In terms of the number of authors who published papers at the conference and workshops (RQ 3), the total number of authors notably decreases for the conference from 363 in the first time period to 282 in the last period -- see Table~\ref{avg_participations}. At the same time, the average number of authors per paper published at the conference grew from 2.9 to 3.3 and 3.5, indicating a greater relevance of collaborations -- see Table~\ref{avg_participations}. The data on affiliations for the conference and workshops suggests considerably fewer publications of authors from research institutions and industry in recent time frames, whereas the number of authors from higher level education institutions has only slightly decreased -- Table~\ref{institution_categories}. In summary, we can observe concentration effects shown by the decreasing number of authors, affiliations, and geography. Further studies would be required to analyze possibly hidden relationships in these areas and reveal the causes for these developments.


The second set of research questions RQ 4 to 6 concerned the content-based analysis of publications with the identification of major topics (RQ 4), differences in academia and industry (RQ 5), and the possible indication of prospective topics (RQ 6). LDA describes topics by individual terms objectively; however, the discussion of topics has a subjective component due to semantic ambiguities. 

For RQ 4, we can characterize each topic's general area by the following subjectively assigned labels, given in decreasing order by cumulative weight -- see Table \ref{LDA topics for all papers from 2005 to 2019 ordered by cumulative topic weight}: These comprise of topics related to (1) 'process mining and event logs', (2) 'business process models and BPMN', (3) 'services and messaging', and (4) 'systems and workflows' as well as (5) 'business, management and research design', (6) 'nets and state-transition', (7) 'data, time and resources', and (8) 'constraints, rules and compliance'. While some topics are relatively general, it has to be noted that LDA assigns papers to multiple topics, thus not limiting the classification. The topics related to process mining and workflows are equally present in the results for the conference and the workshops. 

For the workshops, topics on (4) 'business process models and BPMN' as well as for practical cases such as (8) 'medical, clinical and patient' are more prominent in comparison to the conference -- see Table~\ref{LDA topics for workshop papers from 2005 to 2019 ordered by cumulative topic weight}. For the conference, greater relative importance of (3) '(petri) nets and state-transition' and (4) 'graph-based approaches' can be observed -- see Table \ref{LDA topics for conference papers from 2005 to 2019 ordered by cumulative topic weight}. 

In each LDA analysis of the five-year time frames from 2005 to 2019, changes in the relative importance of topics can be observed -- see Tables~\ref{LDA topics for all papers from 2005 to 2009 ordered by cumulative topic weight}, \ref{LDA topics for all papers from 2010 to 2014 ordered by cumulative topic weight}, \ref{LDA topics for all papers from 2015 to 2019 ordered by cumulative topic weight}. Up until 2009, papers related to the topics (2) 'semantic web, services and ontologies', and (7) 'BPEL' are found. Later, up to 2014, (5) 'object' occurs as a term for the last time and BPEL seems to have lost importance, while (2) 'models, case and management' is introduced, potentially coinciding with the CMMN standard~\cite{omg2014}. Within 2015–2019, (7) 'decision, model, BPMN' occurs first. Despite its comparatively low importance in the time frame 2005-2009, the topic of 'process mining' clearly made its way to the top after that period and until today.


RQ 5 aimed for investigating differences between academia and industry. These were analyzed in Tables~\ref{LDA topics for all papers from 2005 to 2019 with institution category 0 or 1 ordered by cumulative topic weight} and~\ref{LDA topics for all papers from 2005 to 2019 with institution category 2 or 3 ordered by cumulative topic weight}. Here we can observe that the topic (5) 'nets and state-transition' seems to be of higher importance in academia as it does not occur in the top eight topics of the category industry or other papers. On the other hand, (4) 'models and business application', and (8) 'patient and clinical' only occur in the LDA results for papers with industry affiliations, suggesting a focus on execution and implementation topics. Again, in both areas, process mining and workflows seem to be popular.

For answering RQ 6 we can observe prospective topics, application areas, and domains considering the LDA analyses' results. Topics consistently popular in each LDA analysis of the five-year time frames from 2005 to 2019 are process mining, services, workflows, and, (Petri) nets and state-transition. The growth of process mining in the most recent time frame seems to be crowding out other topics. Topics involving the term modeling decreased from 6 topics in 2005-2009 to 5 topics in 2010-2014 to 4 topics in 2015-2019. Specifically, topics related to languages, rules, and graphs are currently on the decrease. In light of the recent establishment of a dedicated process mining conference~\cite{icpm2019}, it could be an opportunity for the BPM conference and workshop community to focus again more on these topics, e.g.\ by engaging more in language development and the role of modeling. Further, tackling the decline in the participation of industry and according topics, e.g.\ regarding application development, could be further opportunities for maintaining the strength of the conference and broaden the community.


When comparing the results of our analysis with previous investigations, we can find the following. In Houy et al.'s automated analysis~\cite{HouySFL13}, a set of BPM terms and concepts was proposed that occur frequently in the abstracts of BPM papers. For comparison, we extracted an exemplary selection of 49 BPM terms and concepts from their result table and matched them with the terms of our LDA analysis. From the three periods 2005-2009, 2010-2014, and 2015-2019, it can be shown that 45\% of the terms from the LDA analysis can be mapped directly to the terms found in~\cite{HouySFL13}. However, the comparison is limited since TF-IDF tends to have a variety of relatively specific subject matters due to the inverse document frequency method favoring terms differentiating topics well, while LDA identifies distributions of frequent topics.


In \cite{neder2018}, 16 BPM themes over time were derived using network centrality measures. Unlike the LDA and TF-IDF, these result from nodes in a semantic network. The individual terms of these network nodes often relate to specific subsets of a topic such as \textit{traditional-BPM}, while our analysis shows individual aspects of the topics such as \textit{BPM}. Therefore, a direct comparison of the results does not seem applicable.

For the analysis in~\cite{VanAalst2013}, our identified LDA topics from 2005 to 2019 suggest the support of four out of six key concerns, identified by the earlier study: \textit{Process Mining}, \textit{Process Modeling Languages}, \textit{Process Enactment Infrastructures} (i.e.\ workflow system topics), and \textit{Process Model Analysis} (i.e.\ data, time and resources). The two key concerns not directly found in our analysis are \textit{Process Flexbility} and \textit{Process Reuse}. When comparing the major topics of our LDA analysis, it however revealed topics that do not seem to be explicitly reflected by the key concerns such as \textit{services}, \textit{business, management and research design}, and \textit{constraints, rules and compliance} - however, this may depend on a subjective view on these topics and would need to be discussed between the authors of each study in detail. In addition, there are methodological differences to our work. The study by~van der Aalst focused on key concerns and use cases, where use cases primarily concern the functions and qualities, not directly capturing topics. For this reason, we only considered key concerns for comparison.


In \cite{ReckerM16}, the phases of the BPM lifecycle were taken as a foundation. The major topics we identified in our LDA analysis from 2005-2019 are consistent with the BPM lifecycle phases \textit{Process discovery} that can be related to the topics of process mining, \textit{Process analysis} and \textit{Process monitoring and controlling} which are related to time, resource and data topics, \textit{Process identification} and \textit{Process re-design} that are related to process models and BPMN as well as business and management topics, and \textit{Process implementation and execution} that is related to transition and workflow topics. In comparison with some key insights of the study by Recker and Mendling, we can confirm that contributions to process improvement and the prominent use of empirical methods that would find their way into the major topics are also absent in our study.


\section{Conclusion, Limitations and Outlook}

In this paper we conducted a computational bibliometric study on the metadata and the contents of papers at the BPM conference and its workshops from 2005-2019. In addition to the obvious limitations of such a study such as the missing manual inspection of the contents of all papers and the limited traceability of the results of the LDA analysis back to individual papers, there are two further aspects that need to be considered when interpreting the results. First, the interpretation of the results of the LDA has only been conducted by the authors and not been reflected yet with others. Second, the study only considered papers of the BPM conference and workshops, which certainly limits the generalizability of the results to BPM as a discipline.
Nevertheless we hope that the results stipulate discussions in the community on the evolution of BPM as a discipline and forum for research. In the future we plan to systematically reflect the gained insights with other members of the community and extend the study to further outlets for enhancing generalizability.


\bibliographystyle{splncs04}
\bibliography{literature.bib}

\end{document}